\documentclass[10pt,a4paper]{article}

\usepackage[dvips]{graphicx}
\usepackage{amsmath}
\usepackage{amssymb}
\usepackage{bm}

\newcommand{\be}{\begin{equation}}
\newcommand{\ee}{\end{equation}}
\newcommand{\curl}{\bm\nabla\wedge}
\newcommand{\divr}{\bm\nabla\cdot}
\newcommand{\calm}{{\mathcal{M}}}
\newcommand{\calq}{{\mathcal{Q}}}

\begin{document}

\title{Diffused vorticity approach to the oscillations\\
of a rotating Bose-Einstein condensate confined\\
in a harmonic plus quartic trap}
\author{M.~Cozzini\\
{\it\small Dipartimento di Fisica, Universit\`a di Trento and BEC centre
CNR-INFM,}\\
{\it\small via Sommarive 14, I-38050 Povo (TN), Italy}}

\maketitle

\abstract{%
The collective modes of a rotating Bose-Einstein condensate confined in an
attractive quadratic plus quartic trap are investigated. Assuming the presence
of a large number of vortices we apply the diffused vorticity approach to the
system.
We then use the sum rule technique for the calculation of collective
frequencies, comparing the results with the numerical solution of the
linearized hydrodynamic equations. Numerical solutions also show the existence
of low-frequency multipole modes which are interpreted as vortex
oscillations.}

\section{Introduction}

Quantized vortices are one of the most striking features of superfluids,
ranging from liquid helium to superconductors. Bose-Einstein condensates of
alkali gases have proved to be one of the best tools to study these fascinating
quantum objects, presenting several advantages with respect to both helium and
superconductors. On the other hand, due to the limited resolution imposed by
current experimental techniques, direct in situ imaging of vortices is a
very difficult task. This problem has been however overcome by using
time-of-flight absorption images, which take advantage of the expansion of the
gas after release of the trap.

In this context, particularly appealing structures are given by vortex arrays,
where singly quantized vortices typically arrange in highly regular triangular
lattices, similar to the Abrikosov lattice of superconductors. In order to
realize such configurations, large amounts of angular momentum have to be
transferred to the gas, what can be done by using various experimental
procedures \cite{exp vort}. The acquired angular velocity tends then to enlarge
the rotating cloud, the centrifugal force giving rise to bulge effects which
flatten the density profile towards a 2D configuration. In the presence of
purely harmonic confinement characterized by a frequency $\omega_\perp$ in the
plane of rotation, this phenomenon fixes an upper limit for the angular
velocity $\Omega$ of the system, namely the frequency $\Omega=\omega_\perp$ at
which the quadratic effective potential given by the centrifugal force exactly
equals the harmonic trapping term. Beyond this angular velocity the centrifugal
force dominates over the confinement and the system is no longer bounded.

The possibility of reaching arbitrarily high angular velocities is however
provided by stronger than quadratic traps \cite{quartic ENS,quartic 3D,quartic
2D}. The introduction of a quartic term in the potential then opens up new
regimes in the study of rotating condensates, making possible the realization
of new equilibrium configurations with different vortex arrays.

In the present contribution, we will first briefly summarize the stationary
solutions for an effectively two-dimensional Bose-Einstein condensate rotating
in a harmonic plus quartic trap at zero temperature \cite{quartic stat}. Then,
within the Thomas-Fermi approximation \cite{RMP}, we will focus on the analysis
of the most important collective excitations  of the system \cite{quartic osc}.
Numerical solution of the hydrodynamic equations will be combined with the sum
rule method.
Vortex arrays will be treated within the so-called diffused vorticity
approximation \cite{stripes}, thereby neglecting the microscopic motion of
single vortices in favour of the macroscopic dynamics of the system.
Nevertheless, signatures of vortex modes will also be found, although the
corresponding predicted frequencies are not expected to be very accurate.
Indeed, a consistent calculation of such effects would require a more detailed
treatment, as the one of Ref.~\cite{B-C}.

\section{Stationary configurations}

As anticipated above, in the presence of a large number of vortices, as always
assumed in the following\footnote{For very large rotation rates the external
potential considered here is predicted to give rise to giant vortex states
\cite{quartic stat}, where all the vorticity is confined in a single hole. We
do not discuss such configurations and always deal with the case where singly
quantized vortices are present.},
it is possible to use the so called diffused vorticity approach, consisting of
averaging the velocity field $\bm{v}$ over regions containing many vortex lines
and assuming that the vorticity is spread continuously in the fluid. For
example, in the case of a uniform vortex lattice with average vortex density
$n_v$, this corresponds to assuming rigid body rotation
$\bm{v}=\bm\Omega\wedge\bm{r}$ with $\Omega=hn_v/2M$, where $M$ is the atomic
mass. The relation between the effective angular velocity $\bm\Omega$, oriented
along the vortex line direction, and the vortex density $n_v$ can be derived by
imposing that the circulation of the averaged velocity field around a single
vortex cell be equal to $h/M$, as for the single vortex case. More generally,
the usual irrotationality condition $\curl\bm{v}=0$ of superfluid hydrodynamics
is broken in favour of $\curl\bm{v}(\bm{r})=hn_v(\bm{r})/M$, where
$n_v(\bm{r})$ is the average vortex density in the proximity of point $\bm{r}$.
It is then clear that the validity condition of this approach is that the
average distance $1/\sqrt{n_v}$ between vortices be much smaller than the size
of the cloud\footnote{For an annular structure one has to compare the average
vortex distance with both the annulus radius and the annulus width.},
the concept of diffused vorticity being adequate to describe the dynamics only
at distances larger than $1/\sqrt{n_v}$. If in addition $\xi\ll1/\sqrt{n_v}$,
where $\xi=\hbar/\sqrt{2gn}$ is the healing length defined in terms of the
interaction coupling constant $g$ and bulk density $n$, one can safely use the
Thomas-Fermi approximation. Indeed, if the size of vortex cores fixed by $\xi$
is much smaller than the inter-vortex distance, one can assume a slowly varying
density profile between vortices, consequently neglecting density gradients
associated with quantum pressure effects.

Within the presently discussed approximations, the system is then described by
the rotational hydrodynamic equations, which in the rotating frame read
\be\label{eq:cont}
\frac{\partial{n}}{\partial t}+\bm\nabla\cdot({n}\bm{v}') = 0 \ ,
\ee
\be\label{eq:euler}
\frac{\partial\bm{v}'}{\partial t}+
\bm\nabla\left(\frac{{v'}^2}{2}+\frac{V_{\text{ext}}}{M}-
\frac{|\bm\Omega\wedge\bm{r}|^2}{2}+\frac{g{n}}{M}\right) = 
\bm{v}'\wedge(\curl\bm{v}')-2\bm\Omega\wedge\bm{v}' \ ,
\ee
where ${n}(\bm{r},t)$ is the spatial density,
$\bm{v}'(\bm{r},t)=\bm{v}(\bm{r},t)-\bm\Omega\wedge\bm{r}$ is the velocity in
the rotating frame, and $V_{\text{ext}}$ is the external potential. It is
trivial to check that $\bm{v}_0=\bm\Omega\wedge\bm{r}$,
$g{n}_0=\mu-V_{\text{ext}}+M|\bm\Omega\wedge\bm{r}|^2/2$ is a stationary
solution for the system, where $\mu$, fixed by the normalization condition
$\int{n}\,\text{d}\bm{r}=N$, is the chemical potential in the rotating frame.

Although equilibrium configurations in Thomas-Fermi approximation can be
obtained also for the 3D case \cite{quartic 3D}, for simplicity we will
consider only 2D configurations, which makes the analysis of collective
excitations considerably easier. In fact, due to the repulsive centrifugal term
$-|\bm\Omega\wedge\bm{r}|^2/2$ in Eq.~(\ref{eq:euler}), which tends to flatten
the equilibrium density ${n}_0$, this is a natural approximation for fast
rotating condensates\footnote{For small angular velocities one needs the
additional assumption that a strong confinement in the axial direction is
present.}. Instead of using the 3D coupling constant $g_{3D}=4\pi\hbar^2a/M$,
where $a$ is the usual 3D $s$-wave scattering length, we then introduce an
effective 2D coupling constant $g_{2D}=g_{3D}/Z$, where $Z$ is a proper length
taking into account the extension of the real system along the rotation
axis\footnote{For a system uniform along the $z$-direction $Z$ corresponds
to the vertical size, while in the case of strong axial harmonic confinement
one has $Z=\sqrt{2\pi}a_z$, where $a_z$ is the oscillator length in the same
direction.}.

The trapping potential is given by
\be\label{eq:ext pot}
V_{\text{ext}} = \frac{\hbar\omega_\perp}{2}\left(\frac{r^2}{d_\perp^2}+
\lambda\,\frac{r^4}{d_\perp^4}\right) \ ,
\ee
where $\omega_\perp$ is the harmonic oscillator frequency,
$d_\perp=\sqrt{\hbar/M\omega_\perp}$ is the characteristic harmonic oscillator
length with the atomic mass $M$, $r=\sqrt{x^2+y^2}$ is the two-dimensional
radial coordinate, and $\lambda$ is the dimensionless parameter characterizing
the strength of the quartic term. In the following, we will use dimensionless
harmonic oscillator units, where $\omega_\perp$ and $d_\perp$ are the units of
frequency and length respectively.

The equation for ${n}_0$ then becomes
\be\label{eq:n0}
{n}_0 =
\frac{1}{g}\left[\mu+\frac{\Omega^2-1}{2}r^2-\frac{\lambda}{2}{r^4}\right] =
\frac{\lambda}{2g}(R_2^2-r^2)(r^2-R_1^2) \ ,
\ee
where $g$ is the dimensionless coupling constant and
\be
R_{1,2}^2 = \frac{\Omega^2-1}{2\lambda}\pm
\sqrt{\left(\frac{\Omega^2-1}{2\lambda}\right)^2+\frac{2\mu}{\lambda}} \ .
\ee
The density is assumed to be zero where the right hand side of
Eq.~(\ref{eq:n0}) is negative. For $\mu>0$ the value of $R_1$ becomes purely
imaginary and the density vanishes at the radius $R=R_2$, while for $\mu\leq0$
two different radii $R_{1,2}$ are present. This reflects the transition
occurring at $\mu=0$, where a hole forms in the centre of the condensate and
the density profile assumes an annular shape. From the normalization condition
one can calculate the transition angular velocity $\Omega_h$ obtaining
$\Omega_h^2=1+(12\lambda^2gN/\pi)^{1/3}$ \cite{quartic stat}.

For the case $\Omega<\Omega_h$, when the hole is absent, expressing $R_1$ in
terms of $R_2=R$ as $R_1^2=(\Omega^2-1)/\lambda-R^2$, the normalization
condition gives the following third degree equation for $R^2$
\be
R^4(4\lambda{R^2}-3\Omega^2+3) = \frac{12gN}{\pi} \ ,
\ee
while the chemical potential becomes $\mu=R^2(\lambda{R^2}-\Omega^2+1)/2$. In
the following we will also need the expectation values
$\langle{r^2}\rangle=\int{n}_0r^2\,\text{d}\bm{r}/N=
\pi{R^6}(3\lambda{R^2}-2\Omega^2+2)/24gN$ and
$\langle{r^4}\rangle=\pi{R^8}(8\lambda{R^2}-5\Omega^2+5)/120gN$.

For the case $\Omega>\Omega_h$, defining $R_{\pm}^2=R_2^2\pm{R_1^2}$ the
normalization condition simply gives
\be
\lambda{R_-^6} = \frac{12gN}{\pi} \ ,
\ee
and hence $R_+^2=(\Omega^2-1)/\lambda$ and $R_-^2=(\Omega_h^2-1)/\lambda$. The
chemical potential is now $\mu=-\lambda(R_+^4-R_-^4)/8$ and the previously
defined expectation values become $\langle{r^2}\rangle=R_+^2/2$ and
$\langle{r^4}\rangle=(5R_+^4+R_-^4)/20$.

\section{Collective modes}

The collective oscillations of the system in the Thomas-Fermi approximation can
be found by linearizing the hydrodynamic equations (\ref{eq:cont}) and
(\ref{eq:euler}), which become
\begin{eqnarray}
&&\frac{\partial}{\partial t}\,\delta{n}+
\bm\nabla\cdot({n}_0\,\delta\bm{v}) \,\ = \,\ 0 \ , \label{eq:HD dn}\\
&&\frac{\partial}{\partial t}\,\delta\bm{v}+g\,
\bm\nabla\,\delta{n}+2\,\bm\Omega\wedge\delta\bm{v} \,\ =
\,\ 0 \ . \label{eq:HD dv}
\end{eqnarray}
These equations can be solved by expressing the radial and azimuthal components
$\delta{v_r}$ and $\delta{v_\phi}$ of the velocity field in terms of
$\delta{n}$ and looking for solutions of the form
$\delta{n}=\delta{n}(r)e^{im\phi}e^{-i\omega t}$, where $m$ is the azimuthal
quantum number, $\phi$ is the azimuthal angle and $\omega$ is the excitation
frequency in the rotating frame. This gives\footnote{It is worth noticing that
not all the solutions of Eq.~(\ref{eq:dn(r)}) correspond to physical density
variations. Indeed, in general one has to check that the resulting
eigenfunctions preserve the density normalization
$\int\,\delta{n}\,\text{d}\bm{r}=0$ and that the corresponding velocity
variations are finite. The latter condition, for example, leads to the
exclusion of the solutions with $\omega=2\Omega$.}
\begin{eqnarray}
&\displaystyle(\omega^2-4\Omega^2)\delta{v_r} =
i\left(-\omega\partial_r+\frac{2m\Omega}{r}\right)g\delta{n}
\ , \label{eq:dv_r}&\\
&\displaystyle(\omega^2-4\Omega^2)\delta{v_\phi} =
\left(-2\Omega\partial_r+\frac{m\omega}{r}\right)g\delta{n}
\ , \label{eq:dv_phi}&\\
&\displaystyle\omega\left[\omega^2-4\Omega^2-\frac{m^2gn_0}{r^2}\right]\delta{n}-
\frac{2m\Omega}{r}\frac{\partial(gn_0)}{\partial{r}}\delta{n}+
\frac{\omega}{r}\frac{\partial}{\partial{r}}
\left(rgn_0\frac{\partial\,\delta{n}}{\partial{r}}\right) = 0 \ .
\label{eq:dn(r)}&
\end{eqnarray}
At first sight, one could expect that the last equation depends both on $gN$
and $\lambda$. Actually, only a given combination of these parameters really
matters: in particular, for given $\Omega$ and $m$, the solution is uniquely
fixed by the value of $\Omega_h$, i.e. by the product $\lambda^2gN$.

Eq.~(\ref{eq:dn(r)}) can be significantly simplified in the large $\Omega$
limit \cite{quartic osc}, where useful analytical results can be obtained. In
general, however, this equation has to be solved numerically, what can be
achieved by direct integration with the natural initial condition\footnote{In
fact, density boundaries are regular singular points of Eq.~(\ref{eq:dn(r)}).
For these points the second derivative term cancels. Notice also that one can
arbitrarily fix the value of $\delta{n}(R_2)$, this choice being equivalent to
imposing the amplitude of the oscillation.}
\be\label{eq:IC}
\frac{\partial\,\delta{n}}{\partial{r}}(R_2) =
\frac{\omega(\omega^2-4\Omega^2)+2m\Omega\lambda(R_2^2-R_1^2)}
{\omega\lambda{R_2}(R_2^2-R_1^2)}\delta{n}(R_2)
\ee
and by varying $\omega$ in order to obtain a well behaved solution. This
procedure can be easily automatized by checking the validity of a condition
similar to Eq.~(\ref{eq:IC}) for the final integration point $r=0$ for
$\Omega<\Omega_h$ and $r=R_1$ for $\Omega>\Omega_h$ and essentially corresponds
to the so called shooting method described in Ref.~\cite{NR}.
The code has also been checked \cite{brian} against the relaxation method
\cite{NR}.

To obtain analytical results also below the large $\Omega$ limit, we will rely
on the sum rule method. To this purpose we introduce the $p$-energy weighted moments
\be\label{eq:moments}
m_p(F) = \sum_n \sigma_n(F) E_{n0}^p
\ee
relative to a generic excitation operator $F=\sum_{k=1}^Nf(\bm{r}_k)$, where
$E_{n0}$ is the energy difference between the excited state $|n\rangle$ and the
ground state $|0\rangle$, and $\sigma_n(F)=|\langle{n|F|0}\rangle|^2$ is the
associated strength. We also define
\be
m_p^\pm(F) = m_p(F) \pm m_p(F^\dagger) \ .
\ee
Notice that for hermitian operators $F=F^\dagger$ one simply has
$m_p^+(F)=2m_p(F)$ and $m_p^-(F)=0$.

The energy weighted moments can be used to derive rigorous upper bounds for the
excitation frequencies of the system \cite{PS book}. For example, one has the
following inequality for the lowest energy $\hbar\omega_{\text{min}}$ excited
by the operator $F$
\be
(\hbar\omega_{\text{min}})^s \leq
\frac{m_{p+s}^+(F)}{m_{p}^+(F)} \ ,
\ee
where $s$ is positive and the equality holds whenever $F$ excites one single
mode.
The explicit calculation of the moments $m_{2p+1}^+(F)$ and $m_{2p}^-(F)$ for
$p\geq0$ can be carried out in terms of commutators between the excitation
operator and the total Hamiltonian of the system, evaluated on the ground
state. In addition, one has the important result $m_{-1}^+(F)=-\chi_F(0)$,
which relates the useful inverse energy weighted moment to the static limit of
the dynamic response function $\chi_F(\omega)$.

We first consider the lowest axisymmetric ($m=0$) mode. One expects that this
breathing oscillation is mainly excited by the monopole operator
$\calm=\sum_{k=1}^Nr_k^2$, although such perturbation, due to the presence of
the quartic term in the potential, slightly couples also to higher
modes\footnote{In the case of purely harmonic trapping the monopole operator is
instead the exact one.}.
Then, as usual, we extract the frequency of the lowest excited mode from the
ratio between the energy weighted ($m_1$) and inverse energy weighted
($m_{-1}$) moments. Indeed, as evident from Eq.~(\ref{eq:moments}), low order
moments minimize the contributions coming from higher
eigenfrequencies\footnote{On the other side, the coupling of the monopole
operator with the next $m=0$ mode can be put in evidence by taking the
$m_3/m_1$ ratio, which is significantly higher than the chosen one.}.
The $m_1$ moment for a hermitian operator can be expressed in terms of
commutators as $m_1(F)=\langle{0|[F,[H,F]]|0}\rangle/2$. Here the many-body
Hamiltonian $H$ in the rotating frame, with an obvious meaning of the symbols,
is given by $H=H_{\text{kin}}+H_{\text{ext}}+H_{\text{int}}-\Omega{L_z}$, where
the interaction term is $H_{\text{int}}=g\sum_{i<j}\delta(\bm{r}_i-\bm{r}_j)$.
For the monopole operator one has $[\calm,[H,\calm]]=2\calm$ and hence
$m_1(\calm)=2N\langle{r^2}\rangle$. On the other hand, since adding a static
monopole perturbation to the Hamiltonian is equivalent to renormalizing the
trapping frequency \cite{PS book}, the monopole static response can be
calculated from $\delta\langle{r^2}\rangle =
\chi_\calm(0)M\delta\omega_\perp^2/2N =
(\partial\langle{r^2}\rangle/\partial\omega_\perp^2)\delta\omega_\perp^2$ (in
dimensional units), where the derivative has to be calculated at constant
angular momentum.
Then, recalling that $m_{-1}(\calm)=-\chi_\calm(0)/2$, in the Thomas-Fermi
approximation one finds
\be \label{eq:m1/m-1 monopole}
\omega^2 = \frac{m_1(\calm)}{m_{-1}(\calm)} =
\left\{
\begin{array}{ll}
6\lambda R^2+4 & (\Omega<\Omega_h) \\
\rule{0cm}{0.6cm}6\lambda R_+^2+4 & (\Omega>\Omega_h) \ ,
\end{array}
\right.
\ee
where the two expressions coincide for $\Omega=\Omega_h$. For
$\Omega>\Omega_h$, since $R_+^2=(R_1^2+R_2^2)=(\Omega^2-1)/\lambda$, one finds
the simple result $\omega=\sqrt{6\Omega^2-2}$.

\begin{figure}
\centerline{\includegraphics[width=8cm]{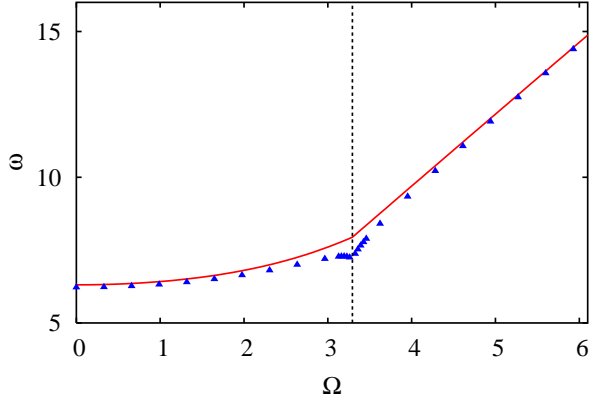}}
\caption{\label{fig:qr M}%
Lowest $m=0$ mode frequency as a function of the angular velocity for
$\lambda=0.5$, $gN=1000$ (frequencies are in harmonic oscillator units). The
solid line is the sum rule estimate, while the triangles are obtained from the
numerical solution of Eq.~(\ref{eq:dn(r)}). The dashed vertical line marks the
critical angular velocity $\Omega_h=3.2935$.}
\end{figure}

Sum rule and hydrodynamic results are reported in Fig.~\ref{fig:qr M}. The
agreement between the sum rule estimate and the numerical solution of
Eq.~(\ref{eq:dn(r)}) is quite good, confirming the hypothesis that the chosen
moments are essentially saturated by the lowest $m=0$ mode. It is worth
noticing that close to the critical angular velocity $\Omega_h$ the
Thomas-Fermi approximation is expected to fail, because quantum pressure
effects become important in the inner region of vanishing density. The sharp
transition shown in the sum rule result and the corresponding dimple in the
hydrodynamic data are in fact smoothed out by the full Gross-Pitaevskii
solution \cite{quartic osc}.

We now switch to excitations of the form $f(\bm{r})=r^{|m|}e^{im\phi}$, which
carry multipolarities different from zero. In particular we will concentrate on
the quadrupole ($m=2$) operator $\calq$. The situation turns out to be much
more complicated than for the monopole operator and one has to include in the
analysis a larger number of moments, treating separately the two regions below
and above $\Omega_h$.

Concerning the case $\Omega<\Omega_h$, one can proceed exactly as in the case
of purely harmonic trapping \cite{zambelli} making a simple 2-mode assumption
and solving the corresponding algebraic system given by the $m_{-1}^+$,
$m_0^-$, $m_1^+$, and $m_2^-$ moments. Indeed one expects that the low order
moments of the quadrupole operator are saturated by the two lowest modes, the
coupling with higher modes being negligible. Since
$m_0^-(\calq)=\langle{0|[\calq^{\dagger},\calq]|0}\rangle=0$, the 2-mode
assumption implies that the strengths of the considered $m=\pm2$ modes are
equal. A simple calculation then shows that the resulting frequencies are
\be
\omega(m=\pm2) =
\frac{1}{2}\left[\sqrt{\left(\frac{m_2^-}{m_1^+}\right)^2+
4\frac{m_1^+}{m_{-1}^+}}\pm\frac{m_2^-}{m_1^+}\right] \ ,
\ee
while for the strengths we have
$\sigma_{m=+2}(\calq)=\sigma_{m=-2}(\calq^\dagger)=m_1^+/[\omega(m=+2)+\omega(m=-2)]$.
For the explicit calculation of the $m_1^+$ and $m_2^-$ moments in the rotating
frame sum rules give
\begin{eqnarray}
m_1^+(\calq) & = & \langle{0|[\calq^{\dagger},[H,\calq]]|0}\rangle =
8N\langle{r^2}\rangle \ , \\
m_2^-(\calq) & = & \langle{0|[[\calq^{\dagger},H],[H,\calq]]|0}\rangle =
-16N(2\Omega\langle{r^2}\rangle-\langle{\ell_z}\rangle) \ ,
\end{eqnarray}
where $\ell_z=-i\partial/\partial\phi$. Since in the Thomas-Fermi diffused
vorticity approach the equilibrium velocity is
$\bm{v}_0=\bm\Omega\wedge\bm{r}$, so that
$\langle{\ell_z}\rangle=\Omega\langle{r^2}\rangle$, one simply finds
$m_2^-/m_1^+=-2\Omega$. In the same approximation, from the static quadrupole
response one has instead $m_{-1}^+=\pi{R^6}/3g$ and hence
$m_1^+/m_{-1}^+=3\lambda{R^2}-2\Omega^2+2$.
Finally
\be\label{eq:om Q nh}
\omega(m=\pm2) = \sqrt{3\lambda{R^2}-\Omega^2+2}\mp\Omega \ .
\ee
At $\Omega=0$ the two lowest $m=\pm2$ modes are degenerate, but, as soon as
some vorticity enters the system, a splitting between the modes
arises\footnote{Notice that at low angular velocities, when only a small number
of vortices is present, the validity conditions of the diffused vorticity
approach are not satisfied, so that Eq.~(\ref{eq:om Q nh}) is only a rough
approximation. At $\Omega=0$, however, it gives the correct Thomas-Fermi
result.}. In the rotating frame, one can then distinguish between a low-lying
and a high-lying branch, with azimuthal quantum number $m=+2$ and $m=-2$
respectively.

\begin{figure}
\centerline{\includegraphics[width=12cm]{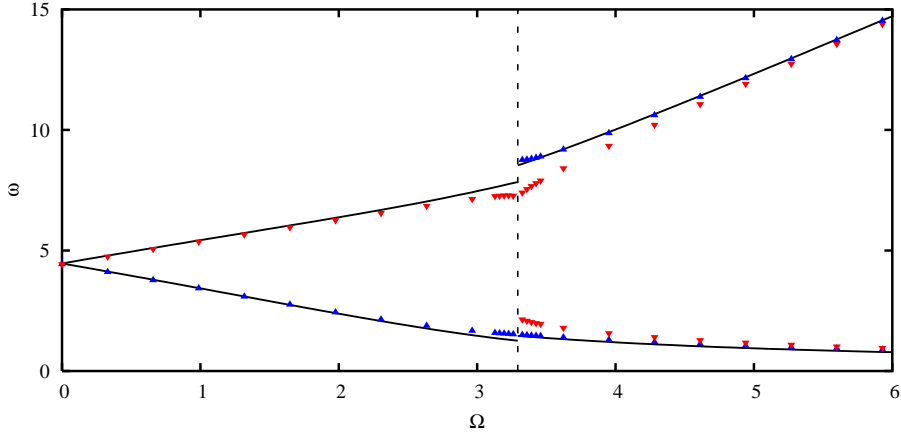}}
\caption{\label{fig:qr Q}%
Excitation frequency as a function of the angular velocity for the main $m=+2$
(upward triangles) and $m=-2$ (downward triangles) modes for the same
parameters as in Fig.~\ref{fig:qr M}. The dashed vertical line corresponds to
$\Omega=\Omega_h$, while the solid lines correspond to the sum rule predictions
discussed in the text.}
\end{figure}

When $\Omega>\Omega_h$, both the low and high-lying branch acquire an
additional mode, as shown by the hydrodynamic numerical results reported in
Fig.~\ref{fig:qr Q}. These new modes, which have opposite azimuthal quantum
numbers with respect to the old ones, do not have any counterpart for
$\Omega<\Omega_h$ and arise due to the annular structure of the condensate.
Their frequencies turn out to be very close to those of their previously
discussed partners, becoming exactly degenerate in the large $\Omega$ limit
\cite{quartic osc}. Hence, since a full treatment of the 4-mode system would be
very complicated, we base our sum rule analysis on the assumption that two
doubly degenerate energy levels are present. In order to calculate the
resulting six unknown quantities $\omega_{\text{H,L}}$,
$\sigma_{\text{H,L}}(\calq)$, and $\sigma_{\text{H,L}}(\calq^\dagger)$, one
needs two more moments, namely
\begin{eqnarray}
m_3^+(\calq) & = &
\langle{0|[[\calq^{\dagger},H],[H,[H,\calq]]]|0}\rangle = \nonumber \\
& = & 16N[(6\Omega^2+1)\langle{r^2}\rangle+
\langle{p^2}\rangle
+3\lambda\langle{r^4}\rangle-
6\Omega\langle{\ell_z}\rangle] \ , \label{eq:m3(Q)}\\
m_4^-(\calq) & = & \langle{0|[[[\calq^{\dagger},H],H],[H,[H,\calq]]]|0}\rangle =
\nonumber \\
& = & -64N[2\Omega(2\Omega^2+1)\langle{r^2}\rangle+2\Omega\langle{p^2}\rangle
-(6\Omega^2+1)\langle{\ell_z}\rangle+ \nonumber \\
&&+3\lambda(2\Omega\langle{r^4}\rangle-\langle{r^2\ell_z}\rangle)]
\ , \label{eq:m4(Q)}
\end{eqnarray}
where $p^2=-(\partial^2/\partial{x}^2+\partial^2/\partial{y}^2)$.
The solution of the corresponding algebraic system gives
\begin{eqnarray}
\omega_{\text{H,L}}^2 & = & \frac{1}{2}\left[\frac{m_4^-}{m_2^-}\pm
\sqrt{\left(\frac{m_4^-}{m_2^-}\right)^2
-4\frac{m_1^+}{m_{-1}^+}\left(\frac{m_4^-}{m_2^-}-
\frac{m_3^+}{m_1^+}\right)}\right] \ , \\
\sigma_{\text{H,L}}(\calq) & = &
\pm\frac{1}{2}\frac{(m_1^+-m_{-1}^+\omega_{L,H}^2)\omega_{\text{H,L}}+m_2^-}
{\omega_{\text{H}}^2-\omega_{\text{L}}^2} \ , \label{eq:s(F)} \\
\sigma_{\text{H,L}}(\calq^\dagger) & = &
\pm\frac{1}{2}\frac{(m_1^+-m_{-1}^+\omega_{L,H}^2)\omega_{\text{H,L}}-m_2^-}
{\omega_{\text{H}}^2-\omega_{\text{L}}^2} \ , \label{eq:s(F+)}
\end{eqnarray}
where, by using the Thomas-Fermi results
$\langle{\ell_z}\rangle=\Omega\langle{r^2}\rangle$,
$\langle{p^2}\rangle=\Omega^2\langle{r^2}\rangle$ and
$m_{-1}^+=\pi(R_2^6-R_1^6)/3g$, one has
$m_1^+/m_{-1}^+=4(\Omega^2-1)\lambda^2R_-^4/[3(\Omega^2-1)^2+\lambda^2R_-^4]$,
$m_3^+/m_1^+=5\Omega^2-1+(3/5)\lambda^2R_-^4/(\Omega^2-1)$ and
$m_4^-/m_2^-=6\Omega^2-2+(6/5)\lambda^2R_-^4/(\Omega^2-1)$.
Recalling that $\lambda{R_-^2}=\Omega_h^2-1$, one can then rewrite the
frequencies as
\begin{eqnarray}
\omega_{\text{H,L}}^2 & = & \displaystyle
3\Omega^2-1+\frac{3}{5}\frac{(\Omega_h^2-1)^2}{\Omega^2-1}+ \nonumber \\
&&\displaystyle\pm\sqrt{\left(3\Omega^2-1+
\frac{3}{5}\frac{(\Omega_h^2-1)^2}{\Omega^2-1}\right)^2
-\frac{4}{5}\frac{5(\Omega^2-1)^2+3(\Omega_h^2-1)^2}{3(\Omega^2-1)^2+
(\Omega_h^2-1)^2}}
\label{eq:om Q wh} \ .
\end{eqnarray}

The results predicted by Eqs.~(\ref{eq:om Q nh}) and (\ref{eq:om Q wh}) are
reported in Fig.~\ref{fig:qr Q}. An analysis of the strengths given by
Eqs.~(\ref{eq:s(F)}) and (\ref{eq:s(F+)}) shows that at $\Omega=\Omega_h$ only
the old modes are significantly excited. However, with increasing $\Omega$,
while the $m=+2$ high-lying mode still has vanishing strength, the $m=-2$
low-lying mode becomes more and more important, eventually even overcoming the
contribution given by the high-lying $m=-2$ mode \cite{quartic osc}. It is also
worth noticing that in the large $\Omega$ limit the high-lying frequency is
essentially given by $\sqrt{m_4^-/m_2^-}$ and one has
$\omega_{\text{H}}^2=6\Omega^2-2$ as for the monopole mode. In the same limit,
the low-lying frequency is given by
$\omega_{\text{L}}=(\sqrt2/3)(\Omega_h^2-1)/\Omega$ \cite{quartic osc}. If one
had used the same 2-mode assumption discussed for $\Omega<\Omega_h$ also for
$\Omega>\Omega_h$, the resulting values would have largely underestimated the
correct frequencies. Finally, we notice that the same procedure could be used
to extract the dipole frequencies.

\begin{figure}
\centerline{%
(a)\includegraphics[width=6cm]{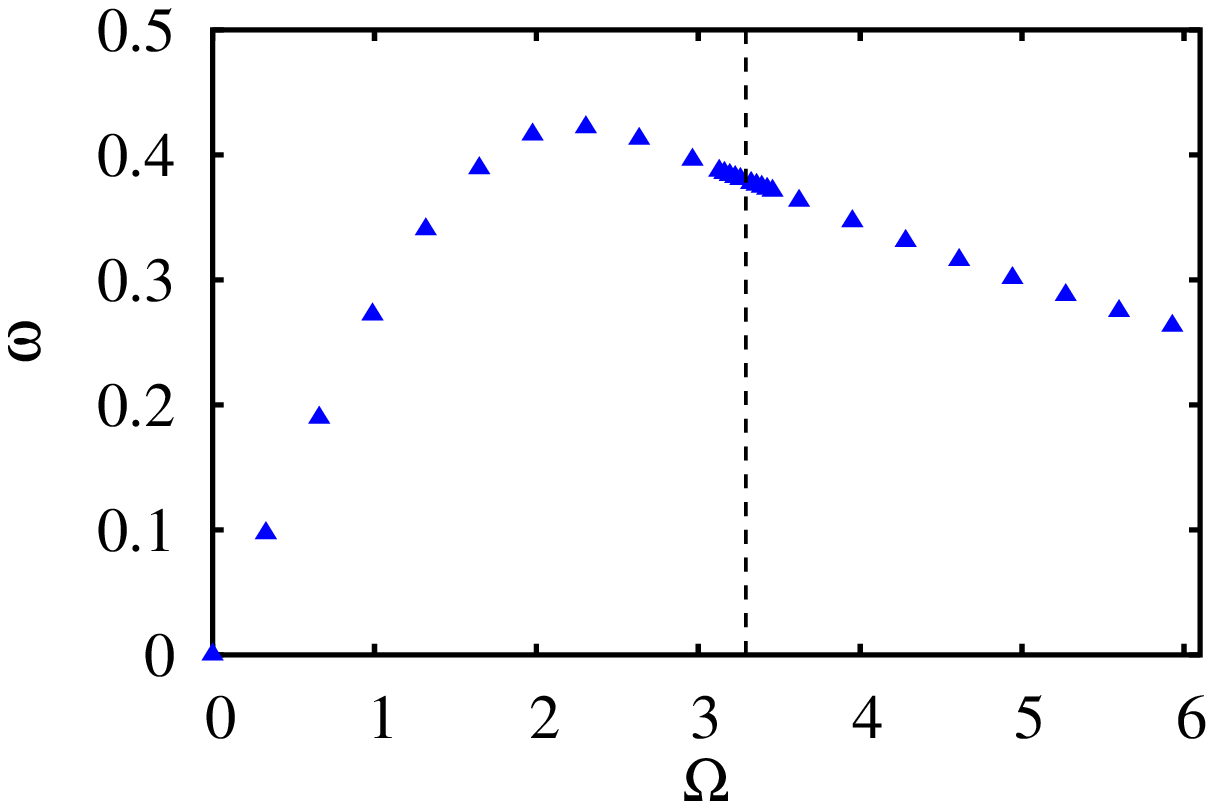}
(b)\includegraphics[width=6cm]{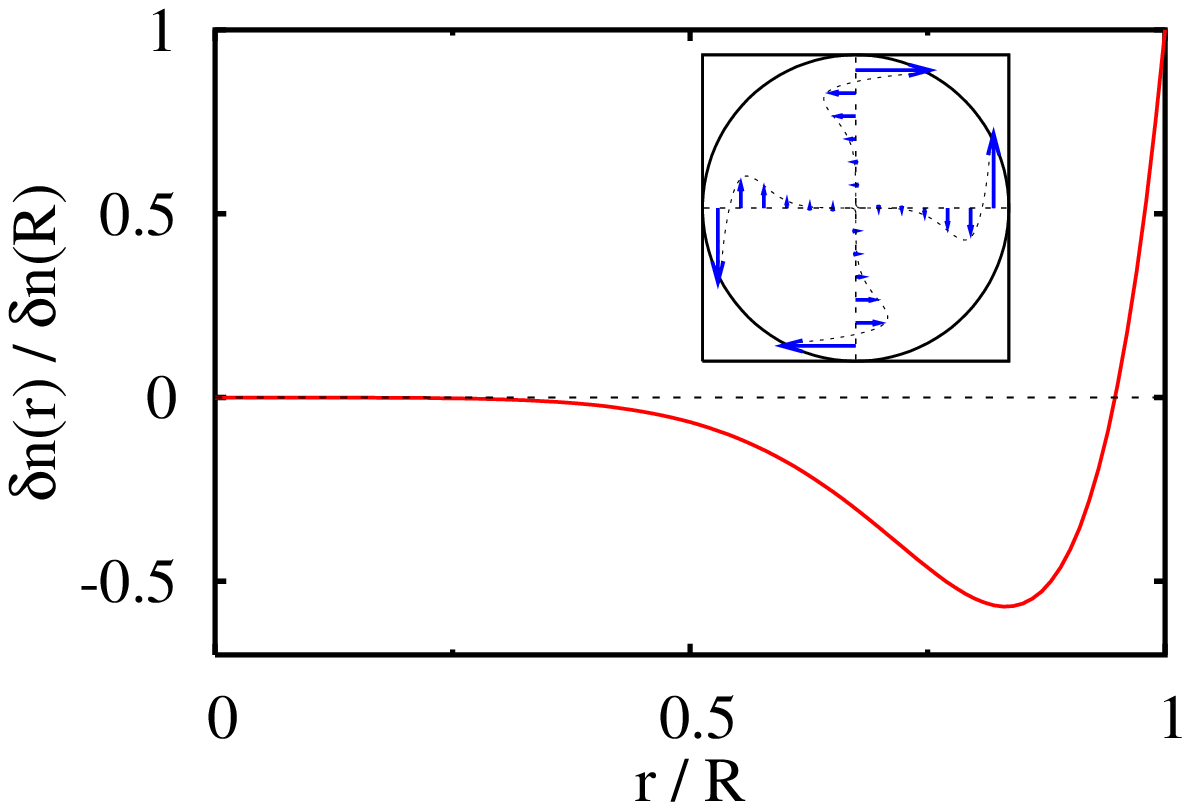}%
\rule[-0.5cm]{0cm}{0.5cm}}
\caption{\label{fig:qr Q tka}%
Behaviour of the $m=+2$ vortex mode with the lowest number of radial nodes for
the same parameters as in Fig.~\ref{fig:qr M}:
(a) excitation frequency $\omega$ as a function of the angular velocity
$\Omega$ (the dashed vertical line corresponds to $\Omega=\Omega_h$);
(b) radial dependence of the density variation at $\Omega=0.9\,\Omega_h$
($\omega=0.396$), the angular dependence being simply given by
${e}^{{i}2\phi}$.
In the inset, the azimuthal component $\delta{v}_\phi$ of the velocity
variation is shown for the angles $\phi=0,\pi/2,\pi,3\pi/2$.}
\end{figure}

In the last part of this section we are going to discuss another class of modes
found from the numerical solution of the hydrodynamic equations. Indeed,
numerical calculations show the presence of very low frequency multipole modes
which cannot be interpreted in terms of sound propagation. One can still label these
excitations with the azimuthal quantum number $m$ and the number of
radial nodes.
The frequencies of the $m=+2$ modes with the lowest number of radial nodes
are plotted in Fig.~\ref{fig:qr Q tka}(a) as a function of the angular velocity.
It emerges that these modes are present on both sides of the critical angular
velocity $\Omega_h$ and that their frequency is zero for $\Omega=0$.
The rotational origin of these excitations together with their very low
frequencies induces to identify these oscillations as multipole vortex modes,
belonging to the family of Tkachenko modes already studied in harmonically
trapped condensates \cite{JILA tka,baym,sr tka,sonin}.
In order to further investigate this hypothesis one can calculate the
corresponding velocity variation, which indeed, at least for $\Omega<\Omega_h$
where the central hole is absent, resembles the typical lattice distortions of
Tkachenko modes. This is shown in the inset of Fig.~\ref{fig:qr Q tka}(b),
where the azimuthal component $\delta{v}_\phi$ of the velocity variation is
plotted, the radial component $\delta{v}_r$ being practically negligible.

Concerning the multipolarity of the modes, a simple remark is in order here. As
noted in Ref.~\cite{chevy}, rotational hydrodynamic equations admit a class of
zero energy solutions which may be interpreted as Tkachenko modes.
More in detail, any zero frequency solution of Eqs.~(\ref{eq:HD dn}) and
(\ref{eq:HD dv}) obeys the relation
\be \label{eq:dv om=0}
\delta\bm{v} = \frac{2\bm\Omega}{\Omega^2}\wedge\bm\nabla{g}\delta{n} \ ,
\ee
which implies $\divr\delta\bm{v}=0$. Hence, substituting into
Eq.~(\ref{eq:HD dn}) and using $\partial{n}_0/\partial\phi=0$, valid for an
axisymmetric equilibrium density profile, one finds
\be
\frac{\partial}{\partial\phi}\delta{n} = 0 \ ,
\ee
so that these modes must correspond to $m=0$. It follows that all the $m\neq0$
modes, included Tkachenko ones, cannot have zero frequency in the diffused
vorticity approach. However, due to the crude approximation used to treat
vorticity within this method, one does not expect the predicted frequencies to
be very accurate.

Actually, it turns out that the numerical frequencies {\it decrease} by
increasing the number of radial nodes.
This is probably related to the fact that the nodes accumulate in the
proximity of the cloud boundary, where the Thomas-Fermi approximation is
expected to fail. This situation is similar to the case of the $m=0$ breathing
mode at the critical angular velocity $\Omega_h$, where the frequencies plotted
in Fig.~\ref{fig:qr M} show an unphysical dimple not present in the full
Gross-Pitaevskii simulations.
It is also worth noticing that the absence of such dimple in the positive $m$
modes of Figs.~\ref{fig:qr Q} and \ref{fig:qr Q tka}(a) is indeed due to the
fact that these oscillations are concentrated on the external boundary (see
Fig.~\ref{fig:qr Q tka}(b)), so that the quality of the approximation used to
treat the central density at $\Omega=\Omega_h$ does not affect their frequency.

According to the proposed picture, the same class of solutions of the rotational hydrodynamic equations must be present also for a purely harmonic potential.
This is indeed the case. For the $m=+2$ mode with the lowest number of radial
nodes, where the predicted frequency is expected to be more reliable, in the
frame rotating at $\Omega=0.7$ one finds $\omega=0.204$ (in units of the
harmonic trapping frequency).
Note that for 2D harmonic trapping the usual multipole modes in the rotating
frame can be found analytically according to the formula $\omega(\pm|m|) =
\sqrt{2|m|-(|m|-1)\Omega^2}\mp\Omega$, identical to the 3D result found in
Ref.~\cite{chevy}.
For the harmonic case, in addition, once the radial distance is expressed in
units of the Thomas-Fermi radius\footnote{In physical units, the Thomas-Fermi
equilibrium density for the 2D harmonic oscillator is
$gn_0=(M/2)(\omega_\perp^2-\Omega^2)R^2(1-r^2/R^2)$.},
Eq.~(\ref{eq:dn(r)}) depends only on $m$ and $\Omega$, so that all the
eigenfrequencies are independent of the interaction.
The diffused vorticity estimate of the multipole Tkachenko frequencies,
consequently, cannot properly include the compressibility effects which have
already proven to be important for the $m=0$ case \cite{sr tka}. Nevertheless,
the comparison with the calculations available in the literature \cite{mizu}
shows that the order of magnitude of the predicted frequencies is correct.

\section{Conclusions}

In this paper we have presented the numerical solution of the two-dimensional
linearized rotational hydrodynamic equations and an accurate sum rule analysis
for the monopole and quadrupole modes of a rotating condensate in a harmonic
plus quartic trap, offering a discussion complementary to the work contained in
Ref.~\cite{quartic osc}.
The numerical results for the monopole and high-lying quadrupole modes have
revealed shortcomings in using the Thomas-Fermi approximation rather than the
full Gross-Pitaevskii solution \cite{quartic osc} near to the critical angular
frequency $\Omega_h$ for hole formation.
The frequency $\Omega_h$ has also been identified as the threshold angular
frequency where additional modes appear, due to the new geometry of the system.
On the other hand, sum rules have provided reliable analytical estimates for
the frequencies and for the excitation strengths of the considered modes.
Finally, it has been shown that the multipole vortex oscillations have a
non-zero energy counterpart in the diffused vorticity approach, in contrast to
the $m=0$ Tkachenko modes. The corresponding frequency estimate is however
expected to be scarcely precise, due to the same effects which lower to zero
the energy in the $m=0$ case.

I warmly thank B.~Jackson, A.L.~Fetter, and S.~Stringari for their precious
suggestions.

\end{document}